# Data-Driven Modeling of Geometry-Adaptive Steady Heat Transfer based on Convolutional Neural Networks: Heat Conduction

Jiang-Zhou Peng[1], Xianglei Liu[2], Nadine Aubry[3], Zhihua Chen[1], Wei-Tao Wu[4*]

1. Key Laboratory of Transient Physics, Nanjing University of Science and Technology, Nanjing, 210094, China

2. School of Energy and Power Engineering, Nanjing University of Aeronautics and Astronautics, Nanjing 210016, China

3. Department of Mechanical Engineering, Tufts University, Medford, MA, 02155, USA

4. School of Mechanical Engineering, Nanjing University of Science and Technology, Nanjing, 210094, China

**Abstract**

Numerical simulation of steady-state heat conduction is common for thermal engineering. The simulation process usually involves mathematical formulation, numerical discretization and iteration of discretized ordinary or partial differential equations depending on complexity of problems. In current work, we develop a data-driven model for extremely fast prediction of steady-state heat conduction of a hot object with arbitrary geometry in a two-dimensional space. Mathematically, the steady-state heat conduction can be described by the Laplace's equation, where a heat (spatial) diffusion term dominates the governing equation. As the intensity of the heat diffusion only depends on the gradient of the temperature field, the temperature distribution of the steady-state heat conduction displays strong features of locality. Therefore, in current approach the data-driven model is developed using convolution neural networks (CNNs), which is good at capturing local features (sub-invariant) thus can be treated as numerical discretization in some sense. Furthermore, in our model, a signed distance function (SDF) is proposed to represent the geometry of the problem, which contains more information compared to a binary image. For the training datasets, the hot objects are consisting of five simple geometries: triangles, quadrilaterals, pentagons, hexagons and dodecagons. All the geometries are different in size, shape, orientation and location. After training, the data-driven network model is able to accurately predict steady-state heat conduction of hot objects with complex geometries which has never been seen by the network model; and the prediction speed is three to four orders faster than numerical simulation. According to the outstanding performance of the network model, it is hoped that this approach can serve as a valuable tool for applications of engineering optimization and design in future.

(The code will be available upon the publication of the manuscript: https://github.com/njustwulab)

| Nomenclature | | | |
|---|---|---|---|
| $a_l$ | $l^{th}$ layer feature map matrix | $S$ | kernel stride |
| $b_l$ | $l^{th}$ layer bias vector | $T$ | temperature field by CFD |
| Conv | convolution layer | $\hat{T}$ | temperature field by ROM |
| Deconv | deconvolution layer | $v_t$ | second order moment |
| $D(X)$ | oriented distance function | $w$ | width of convolution kernel |
| $err$ | average relative error | $W_l$ | $l^{th}$ layer weights matrix |
| $E (\%)$ | deviation | $x$ | x-coordinates |
| $F$ | convolution kernel | $X$ | position vector |
| $g_t$ | gradient of loss function | $y$ | y-coordinates |
| $h$ | height of convolution kernel | $Z$ | zero-level set |
| $H_l$ | size of $l^{th}$ layer feature map | *Greek symbols* | |
| $J$ | loss function | | |
| $l_n$ | layer number | α | learning rate |
| $L$ | relative distance | $\beta_1, \beta_2$ | exponential decay rate |
| $L_r$ | reference length | $\epsilon$ | constant term |
| $m_t$ | first order moment | δ | Mask function |
| $n_F$ | number of convolution kernel | $\theta$ | neural network parameters |
| $N$ | number of datasets | λ | regularization coefficient |
| $N_1, N_2$ | size of output after convolution | σ | RELU activation function |
| $P$ | padding size | $\phi$ | continuous level set function |
| $R^2$ | two-dimensional real space | Ω | computational domain |
| $sign$ | sign function | | |

## 1. Introduction

The steady-state heat equation without a heat source can be described by the Laplace's equation. For complex two- or three-dimensional problems, Laplace's equation must be solved numerically. In general, numerically solving Laplace's equation is not computationally expensive, but this no longer be true for optimization problem which usually requires many times of simulation. During early stage of design/optimization, multiple design alternatives should be explored, and evaluated iteratively. At this stage, it usually does not require high-fidelity simulations, but it is favored that designer can get immediate or real-time feedback for quick iteration. One of the effective alternatives to these high-fidelity numerical heat transfer is reduced order model (ROM), which is computationally efficient [1]–[3].

Reduced order modeling is a consistent research topic for many areas, including heat transfer. Petit et al. first proposed the modal identification method for reduced order modelling of a heat conduction problem [4]. The authors demonstrated the possibilities for a multi-input, multi-output, linear system to obtain a reliable reduced model by outlining identified eigenmodes related to each input. In other hand, a fast and efficient method based on the proper orthogonal decomposition (POD) provides powerful tools building ROM for heat transfer problems [5]–[7]. Ding et al. [8] examine the feasibility and efficiency of the POD based algorithm for developing ROM of heat transfer. Their proposed algorithm can reduce the computation time while keeping high accurate prediction. However, the POD method is also limited because it is a linear combination of eigenvectors and eigenvalues and does not explicitly account for the nonlinear interactions of the highly nonlinear dynamic system [9].

In recently years, the explosion of machine learning (or deep learning, DL) powered methods has been showing power on dealing with reduced order modelling of highly nonlinear dynamic problems. The big difference between the deep learning based and traditional ROM (such as POD, DMD) is that the deep learning uses a stack of simple functions to approximate complicated differential equations, and can be seen as a discretization of the continuous dynamical systems [10]. It should be noticed that most problems in physical sciences are modeled using dynamical systems in the form of differential equations; the continuous dynamical systems can be integrated as a module in deep neural networks for end-to-end learning which makes it simple to building combining between machine learning and physical modeling [11]. The deep learning enable ROM has been proved to be able to accurately capture the spatial and temporal nonlinear features of fluid dynamics system [12]. For instance, Sekar et al. [13] proposed a data driven model by applying deep learning for predicting the steady laminar flow over an airfoil. Their model successfully built a mapping between airfoil geometry, Reynolds number and attack angle, and steady flow fields. Jin et al. [14]. proposed a deep learning reduce order model to predict the unsteady velocity field around a circular cylinder. Their network established a direct mapping between pressure distribution on the cylinder and velocity flow fields.

According to literature review, except the reduced order modelling, recently machine learning approach also has been used to improve the "constitutive modelling" of complex heat flux. Such a data-driven modeling method relieves the inaccuracy caused by assumptions (such as isotropic thermal diffusivity) made during model development or derivation, which leads to a more accurate prediction. Sotgiu et al. [15] proposed a machine learning based method for modeling the model coefficients, and their framework can accurately predict the turbulent heat fluxes. Scalabrin et al. [16] applied artificial neural network (ANN) model to study the heat transfer coefficient during boiling process. The accuracy of the ANN model is much higher than five conventional correlations. Zhou et al. [17] employed the extract learning machine (ELM) algorithm to extract relationship between temperature and air velocity in an simplified air-conditioned room. The trained ELM model can be used to better serve the study of Heating, ventilating and air conditioning (HVAC) system control. Kwon et al. [18] used Random forests algorithm to predict the convection heat transfer coefficients for a high-order nonlinear heat transfer problem. Krishnayatra et al. [19] used k-Nearest Neighbor algorithm as a regression tool to predict the thermal performance of fins. Wei et al. [20] applied the machine learning method to predict the effective thermal conductivity of composite materials.

The deep learning (DL) based reduced order modeling has also begun to cause the attention of thermal engineering community, but so far there are only a few of publications available. In 2018, using convolutional neural network (CNN) Sharma et al. [21] developed a ROM building mapping between boundary conditions and temperature field. Their study demonstrates that it is possible to use DL-ROM to speed up the solution of the Laplace equation; that is learning the physics directly from data. Gao et al. [22] proposed a physics-constrained CNN learning architecture to solve PDEs on irregular domains, where the PDEs includes heat equation and steady Navier-Stokes equations. Their network models achieve a good accuracy of prediction for temperature field and velocity field. In current study, we apply CNN to build a reduced-order, geometry-adaptive heat conduction model, since CNN has demonstrated strong feasibility in geometry representation and per-pixel prediction in images (fields) [23]–[25]. Furthermore, the reduced order model we designed in this paper requires a large datasets to train, therefore memory requirement is an important factor to consider during the model development, while the sparse connectivity and weight-sharing property of CNN reduce the memory cost greatly [26]. The network model constructs a mapping between signed distance function representing the geometry and the temperature field. The paper is organized as follows. In section 2, we introduce the methods, including the network architecture of the reduced order model, the preparation of the datasets and the training algorithm. In section 3 and 4, the results of the cases with complex geometry predicted by the trained network model are presented and discussed. Finally, in section 5, we summarize current work and outlook some potential directions for future exploration.

## 2. Methods

In this paper, we propose a data-driven reduced order model based on deep convolutional neural networks (CNNs). The proposed network model establishes a mapping between temperature field and signed distance function (SDF) which represents the geometry of the problem studied. The network architecture and its training are implemented using Tensorflow. The training datasets are generated by numerically solving the Laplace's equation based on OpenFOAM.

### 2.1. Design of neural network model

#### 2.1.1. Model workflow

Our approach has two key components. Fig. 1 depicts the overall graph of the proposed reduced-order model, which shows the learning strategy and the method of a new prediction of the network model. For the learning/training of the model, we use signed distance function (SDF) which represents the geometry of the problems studied as the network input, and the results (temperature field) of numerical simulation as the learning target (output/label) of the CNNs. After training with proper amount of dataset, the trained model could predict the temperature field based on the new SDF matrix as the input.

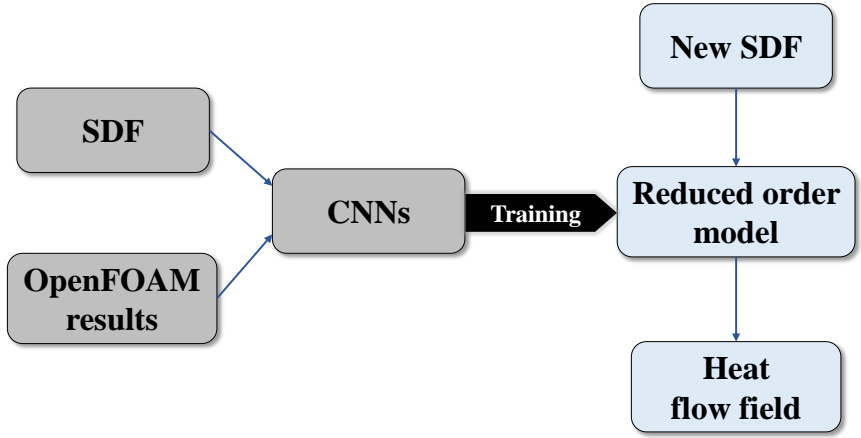

Fig. 1 Schematic of CNNs based reduced order model.

#### 2.1.2. Signed distance function

In this paper, we use a signed distance function (SDF) as the method of geometry representation. Using the SDF, the geometry of the object is represented as a level set function defined over the studied/simulated space in which the object is embedded [27], [28]. Compared to the binary image, the SDF provides more physical and mathematical information for CNNs.

In order to compute the SDF, we need to first create the zero-level set, which represents the boundary of the object. That is, the boundary of the object, $Z$, is represented as the zero-level set of a continuous level set function, $\phi$, defined in a domain $\Omega \subset R^2$; i.e.,

$$Z = \{X \in R^2 : \phi(X) = 0\} \tag{1}$$

The level set function $\phi(X)$ is defined everywhere in the domain $\Omega$, and $\phi(X) = 0$ if and only if $X(x_i, y_i)$ is on the object boundary, such as the edge of the polygon shown in Fig. 2. The SDF associated to a level set function $\phi(X)$ is defined as,

$$D(X) = \min_{X_2 \in \Sigma} |X(x_i, y_i) - X_2(x', y')| sign(\phi(X)) \tag{2}$$

$D(X)$ is an oriented distance function, and the sign function "$sign$" is defined as:

$$sign(x) = \begin{cases} 1 & if\ x > 0 \\ 0 & if\ x = 0 \\ -1 & if\ x < 0 \end{cases} \tag{3}$$

This approach can be used for very accurate calculations if the location of the object boundary is accurately known [29]. Fig. 2 gives a typical example of the SDF representation.

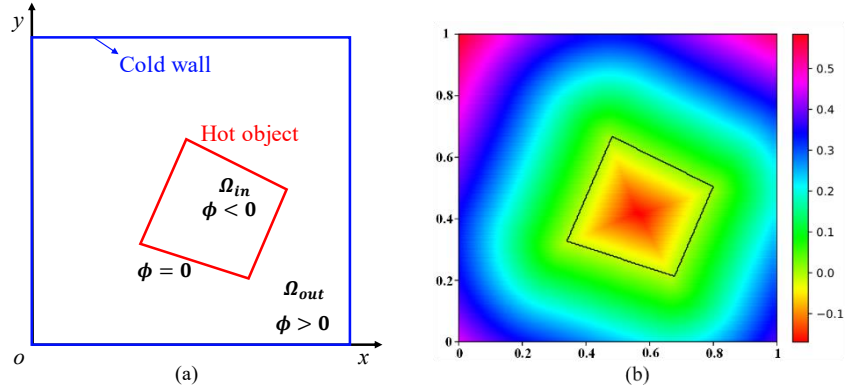

Fig. 2 (a) Schematic of the object and the studied domain; (b) Distribution of the SDF representation of a quadrangle, where the boundary of the quadrangle is in black, and the magnitude of the SDF values equals to the minimal distance of the space point to the quadrangle boundary.

### 2.1.3. Architecture of CNNs

The proposed network architecture consists of two main parts, the encoding part and decoding part. We use a multilayer encoding and decoding architecture for our neural network based reduced-order model. In the model, multiple convolutional layers are applied to extract a highly encoded [30] geometry representation from the SDF (input matrix of the network), and the encoded geometry representation is decoded by multiple deconvolutional layers [31] to generate the temperature field. Here, we use the transpose convolution, called deconvolution, to construct the decoding layers with multiple stacking. Such a structure has been applied to predict the fluid flow field. The operation of these two layers/parts of the model will be briefly described in the next two paragraphs, respectively. Fig. 3 shows the structure and components of the CNNs model, and Table 1 displays the parameters of each layer of the network model.

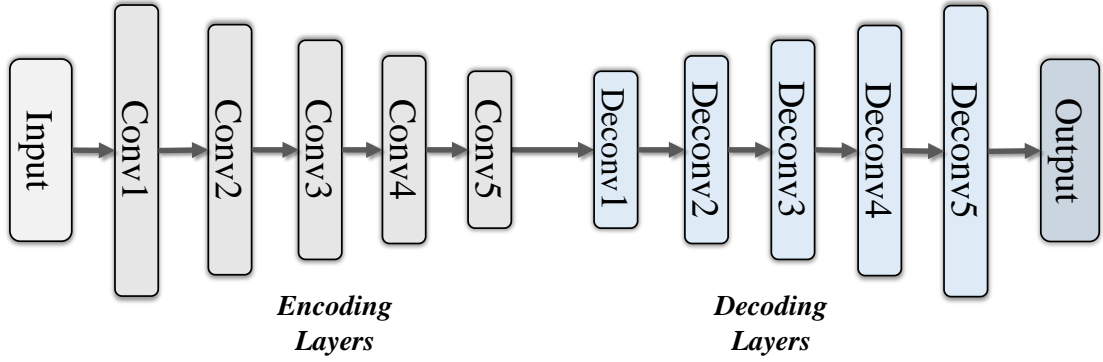

Fig. 3 Architecture of the CNNs model. 'Conv' denotes the convolutional layer; 'Deconv' denotes the deconvolution layer.

Table 1 Parameters of the convolutional layer encoder, where $w \times h$ denotes the size of convolution kernel ($F$); $n_F$ denotes the number of $F$; $N_1 \times N_2 \times n_F$ is the size of the output after convolution.

| Name of Layer | Executing operation $w \times h \times n_F$/stride | Shape $N_1 \times N_2 \times n_F$ |
|---|---|---|
| **Input of model** | -- | 250×250×1 |
| **Conv1** | 5×5×32/5 | 50×50×32 |
| **Conv2** | 4×4×64/2 | 24×24×64 |
| **Conv3** | 4×4×128/2 | 11×11×128 |
| **Conv4** | 2×2×256/1 | 10×10×256 |
| **Conv5** | 2×2×512/1 | 9×9×512 |
| **Deonv1** | 2×2×256/1 | 10×10×256 |
| **Deconv2** | 2×2×128/2 | 20×20×128 |
| **Deconv3** | 3×3×64/2 | 41×41×64 |
| **Deconv4** | 3×3×32/2 | 83×83×32 |
| **Deconv5** | 4×4×1/3 | 250×250×1 |
| **Output** | Reshape | 250×250 |

Mathematically the convolution operation can be expressed as, $\boldsymbol{W}_l * \boldsymbol{a}_{l-1}$, where $\boldsymbol{W}_l$ is the weights or convolutional kernel of the current layer, $\boldsymbol{a}_{l-1}$ is the input and the output of the current and the last layer respectively, and $*$ is the convolutional operator. The schematic of the convolution operation is shown in Fig. 4, where the $4 \times 4$ white matrix denotes the convolutional kernel, the gray matrix is the input matrix and the dark gray matrix represents the output of the convolution operation. Besides the size of the kernel, $F(w \times h)$, the convolutional output is also influenced by the kernel stride, $S$, and the padding size, $P$. The padding operation adds zeros around the border of the input matrix, and the kernel stride controls the sliding step size of the kernel. The size of the output matrix after the convolutional operation can be calculated as [32],

$$H_{l+1} = \frac{H_l - F + 2P}{S} + 1 \qquad (4)$$

where $H_l$ is the size of feature map at $l^{th}$ layer. In current work, zero padding size ($P$=0) is used. From the above equation, it can be seen that after several convolutional operations, the size of the original input can be reduced significantly, and the features of the original input is also highly encoded, thus the memory space required by CNNs training was reduced.

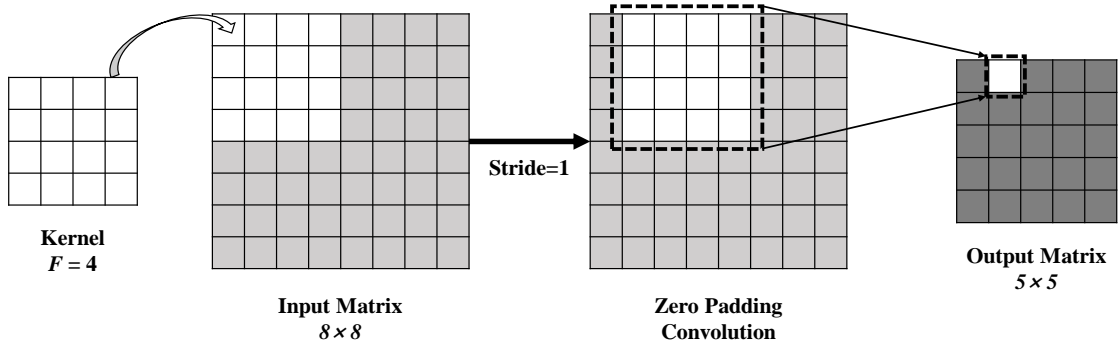

Fig. 4 Schematic of the convolution operation. The white matrix denotes a 4×4 convolutional kernel, and the gray matrix represents the input feature, the dark gray matrix represents the output of convolution operation.

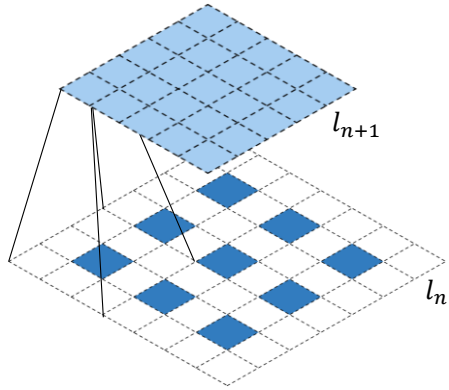

Fig. 5 Schematic of the deconvolutional layer: the dark blue block represents the feature map at the $l_n$ layer, the white block illustrates padding location of $l_n$ layer's feature map, the pool blue block represents the feature map at next layer ($l_{n+1}$).

Fig. 5 shows the schematic of the deconvolution operation. Decoding part is designed to analyze the highly encoded features and decode them to be a recognizable temperature field. Deconvolution operation needs appropriate basis functions to learn invariant subspaces of feature map, and then compensates the minutia information to predict the temperature field by multilayer iterative optimization. For instance, as shown in Fig. 5, for adjusting the width and height of the output feature map (pool blue block), random values (optimized by the algorithm) will be padded at the white block around the block (dark blue block) of the input feature map. Therefore, by applying the deconvolutional decoder, highly encoded information is recovered. Meanwhile, the CNNs model learns/trains its ability of predicting temperature field based on the SDF through this process.

Finally, it is worth mentioning that to increase the non-linear capability of CNNs each convolution or deconvolution layer involves the nonlinear activation operations. The nonlinear activation can be expressed mathematically as,

$$\boldsymbol{a}_l = \sigma(\boldsymbol{W}_l * \boldsymbol{a}_{l-1} + \boldsymbol{b}_l) \qquad (5)$$

where $\boldsymbol{a}_l$ is the output of the $l^{th}$ layer, $\sigma$ denotes the nonlinear activation function, and $\boldsymbol{b}_l$ is the bias. The introduction of $\sigma$ is crucial for neural networks to possess non-linearity. In this paper, we apply the Rectified linear unit (RELU) activation function,

$$\sigma(x) = \begin{cases} 0, x < 0 \\ x, x \geq 0 \end{cases} \tag{6}$$

The advantages of RELU are that its computation cost is cheap as the function has no complicated math and it converges fast thus the model takes less time to train or run. Most important, RELU is sparsely activated [33], and a sparse deep neural network model usually has better prediction performance and less overfitting/noise problem. In current work, the region with negative value of the SDF, which is inside of the object, will not be activated, and this is meaningful physically. Further notice that, because the final prediction of the network model needs to be a continuous regression, therefore between the output layer and the 'Deconv5' layer there is no activation function.

**2.2. Model training**

The model training is an iterative process, which continuously minimizes the loss between the predicted ($\hat{T}$) and the target ($T$) outputs, for obtaining the optimal model parameters, $\boldsymbol{\theta}$. In general, the loss function is defined as,

$$J = \frac{1}{N}\sum_{n=1}^{N}\left(T(s_n) - \hat{T}(s_n)\right)^2 + \lambda||W||_2 \tag{7}$$

where $s_n$ denotes the $n$-th matrix of SDF values, $T(s_n)$ indicates the result by the numerical simulation, $\hat{T}(s_n)$ is the result predicted by the network model, $W$ denotes all the weight of the network layers, $\lambda$ is the regularization coefficient, and $\lambda||W||_2$ is the L2 Regularization term for preventing model overfitting. It should be noticed that inside the hot object there is no physical field. One can map the physical domain outside the hot object to be a structured shape but will be difficult for the objects with complex geometry. We consider two ways to deal with this. First method assumes that the hot object can be considered as a lumped system, therefore the temperature inside it is same to its boundary, and the loss function is calculated by equation (7). Second method ignores the loss attributed to the inside of the hot object, by modifying the loss function to be,

$$J' = \frac{1}{N}\sum_{n=1}^{N}\left(\left(T(s_n) - \hat{T}(s_n)\right)\cdot\delta(s)\right)^2 + \lambda||W||_2 \tag{8}$$

where $\delta(s)$ is defined as,

$$\delta(s) = \begin{cases} 1, \phi(X) \geq 0 \\ 0, \phi(X) < 0 \end{cases} \tag{9}$$

We call the loss function calculated by first method and second method as Loss Function 1 and Loss Function 2.

The backpropagation method with Adaptive Moment Estimation (Adam) method is implemented as the optimization algorithm. With Adam, the model parameters vector, $\boldsymbol{\theta}$, is iteratively updated through:

$$\boldsymbol{\theta}_{t+1} = \boldsymbol{\theta}_t - \alpha \frac{m_t^{corrected}}{\sqrt{v_t^{corrected}} + \epsilon} \tag{10}$$

where $\alpha$ is the learning rate and $\epsilon$ is a number to avoid singularity. $m_t$ and $v_t$ denotes the first and second order moments of the gradients of the loss function [17]-[18], respectively.

$$\begin{aligned} m_t &= \beta_1 m_{t-1} + (1-\beta_1)g_t \\ v_t &= \beta_2 v_{t-1} + (1-\beta_2)g_t^2 \end{aligned} \tag{11}$$

where $\beta_1$ and $\beta_2$ are the exponential decay rates for the moment estimates, $g_t$ denotes the gradients of the loss function with respect to the parameters of the network, $m_t$ and $v_t$ are the estimates of the first moment (the mean) and the second moment (the variance). These two moments control the update direction of the model weight and the step length (learning rate), respectively. Since $m_t$ and $v_t$ are initialized using zero vectors, from above equations Adam optimizer is biased towards zero, especially during the initial time steps, or when the learning rate is small. Therefore, the following correction is proposed to counteract these biases (Note that the superscript $t$ of $\beta$ is presented exponentially),

$$m_t^{corrected} = \frac{m_t}{1 - \beta_1^t}$$
$$v_t^{corrected} = \frac{v_t}{1 - \beta_2^t}$$
(12)

The hyper-parameters of the optimization algorithm for the network training are shown in Table 2. To improve the computational efficiency and the quality of the model, a minibatch-based learning strategy is used during the optimization procedure [36].

Table 2 Hyper-parameters of the optimization algorithm.

| Hyper-parameter | Value |
| --- | --- |
| Batch size | 64 |
| $\lambda$ | 0.00002 |
| $\beta_1$ | 0.99 |
| $\beta_2$ | 0.999 |
| $\epsilon$ | 0.0001 |
| $\alpha$ | 0.0001 |

### 2.3. Preparation of dataset

A good dataset is an important aspect for deep learning to achieve better training and prediction performance. Lack of enough training data, which has nothing to do with the quality of the network structure, can endanger the accuracy of the network model. In this section, we introduce dataset generation and configuration of the numerical solver.

### 2.3.1. Data and preprocessing

In current work, the training and test datasets consist of 5 types of 2D simple objects in a simulated domain: triangles, quadrilaterals, pentagons, hexagons and dodecagons; see Fig. 2 as an example. With the simulated domain kept constant, each set of the objects are randomly different in size, shape, orientation and location. In general, it is more often that the object locates near the center of the studied domain, therefore the location of the object is generated following a normal distribution in the function of $x$ and $y$ coordinates and using center of the domain as the mean. The training and testing dataset contains 5,000 samples (1,000 random samples for each object), and the test dataset contains 750 samples (random separated from the whole training and testing dataset). The simulated domain is discretized using an unstructured mesh tool, SnappyHexMesh [37], [38]. Then the numerical simulation is performed based on the library of OpenFOAM, where SIMPLE algorithm is used. Regarding validation dataset, which is dedicated to validating and also indicating the generalization ability of our CNNs model, more complex geometries of the objects are chosen, see Fig. 6. Those samples have never been seen by the network model during the training process. All the process of the mesh generation and numerical simulation for validation dataset are same to the training dataset.

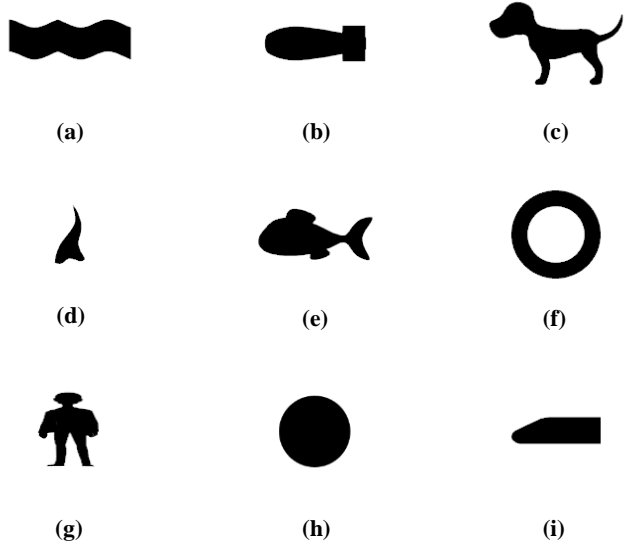

Fig. 6 Geometry of the objects of the validation dataset, (a) Bend, (b) Bomb, (c) Dog, (d) Fire, (e) Fish, (f) Loop (g) Human, (h) Circle, (i) Locomotive.

### 2.3.2. Governing equations

The physics of the studied steady-state heat conduction problem can be mathematically described by Laplace's equation, which is,

$$0 = \nabla \cdot (\nabla T) \tag{13}$$

where $T$ is the temperature, $\nabla \cdot$ is the divergence operator, and $\nabla$ is the gradient operator. The above equation can be normalized as,

$$0 = \nabla^* \cdot (\nabla^* T^*) \tag{14}$$

where we have used the following non-dimensional parameters,

$$T^* = \frac{T - T_0}{T_1 - T_0}; \tag{15}$$
$$\nabla^* \cdot = L_r \nabla \cdot, \nabla^* = L_r \nabla$$

where $L_r$ is a reference length, $T_0$ is the temperature of the cold wall and $T_1$ is the temperature of the hot object. The above equation is numerically solved using OpenFOAM. In the following, the asterisk of $T^*$ will be ignored for simplicity.

### 3. Results

We use the average relative error to measure the prediction accuracy of the network model,

$$err^n = \frac{\sum_x \sum_y err^n_{(x,y)} \cdot \delta_{(x,y)}(s)}{\sum_x \sum_y \delta_{(x,y)}(s)} \tag{16}$$

where $(x, y)$ is the index of the space point, and $err^n_{(x,y)}$ is defined as,

$$err^n_{(x,y)} = \frac{|T_{(x,y)}(s_n) - \hat{T}_{(x,y)}(s_n)|}{T_{(x,y)}(s_n)} \tag{17}$$

From equation (16), it should be noticed that when calculating the error, we only consider the domain outside the hot object.

**3.1 Advantage of SDF Representation**

In current work, for the neural network inputs we used SDF as the geometric representation; while in most previous CNNs based reduced order model a binary image is used instead. To evaluate the effectiveness of the SDF, with the exact same network architecture we also trained a model using binary image as the geometric representation. For a binary image, its element is 0 if and only if the position is on the object boundary or inside the object. The prediction accuracy of different models using different loss functions on test dataset and validation dataset is summarized in the Table 3.

Table 3 Test and Validation prediction errors of SDF and binary representations

| Datasets | Loss function | SDF | Binary |
| --- | --- | --- | --- |
| Test | Loss 1 | 5.49% | 6.59% |
|  | Loss 2 | 1.72% | 6.52% |
| Validation | Loss 1 | 37.66% | 80.02% |
|  | Loss 2 | 5.56% | 53.1% |

With the same hyper-parameters and architecture of deep neural network, the above result indicates that the SDF representation is much more effective than the binary representation. Regarding the validation dataset, the error of using the SDF representation is significantly smaller than the error of using the binary representation. Each value in the SDF matrix carries a certain level of global information, while the values in binary matrix only carry information on the boundary of the object. Therefore, the SDF representation achieves a much better performance than the binary representation. If we want to improve the accuracy of the binary representations, the neural network may require more deep architectures to capture whole complex geometric information properly. However, it means more computational resources will be needed. We also compare the performance of the network model using different loss functions defined in Section 2.2. We can see that Loss Function 2, which applies the constrain with $\delta(s)$, gives a much better performance. In the following investigation, we will apply the SDF representation with Loss Function 2.

**3.2 Performance of the network model**

The main features/advantages of the reduced order model we care about are its ability of accurate and fast prediction on the physical field, which is steady-state temperature distribution here. We first check the accuracy of the network model. Fig. 7 shows two typical temperature fields of the test dataset predicted by the network model and numerical simulation (OpenFOAM), and the corresponding relative error distribution. After proper model training, the prediction accuracy on the test dataset approaches to be higher than 98.3%. Fig. 8 shows the temperature fields of the validation dataset predicted by the network model and numerical simulation (OpenFOAM), and the corresponding relative error distribution. For the four cases shown in Fig. 8, the overall accuracy is higher than 95.72%, and the relative results show that the large error happens frequently near the hot objects. Considering that during the training process, the network model has only seen triangles, quadrilaterals, pentagons, hexagons and dodecagons, such a high prediction accuracy on the validation cases with much complex geometries indicates a strong extensionality of the network model. Fig. 9 shows temperature profiles along *x*-direction at $y_1 = 125; y_2 = 175$, by the network model (lines) and numerical simulation (symbols), and it quantitatively affirms the accuracy and validation of the network model. In other words, these results illustrate the CNNs based reduced-order model is able to predict the temperature field accurately as an alternative to numerical simulation.

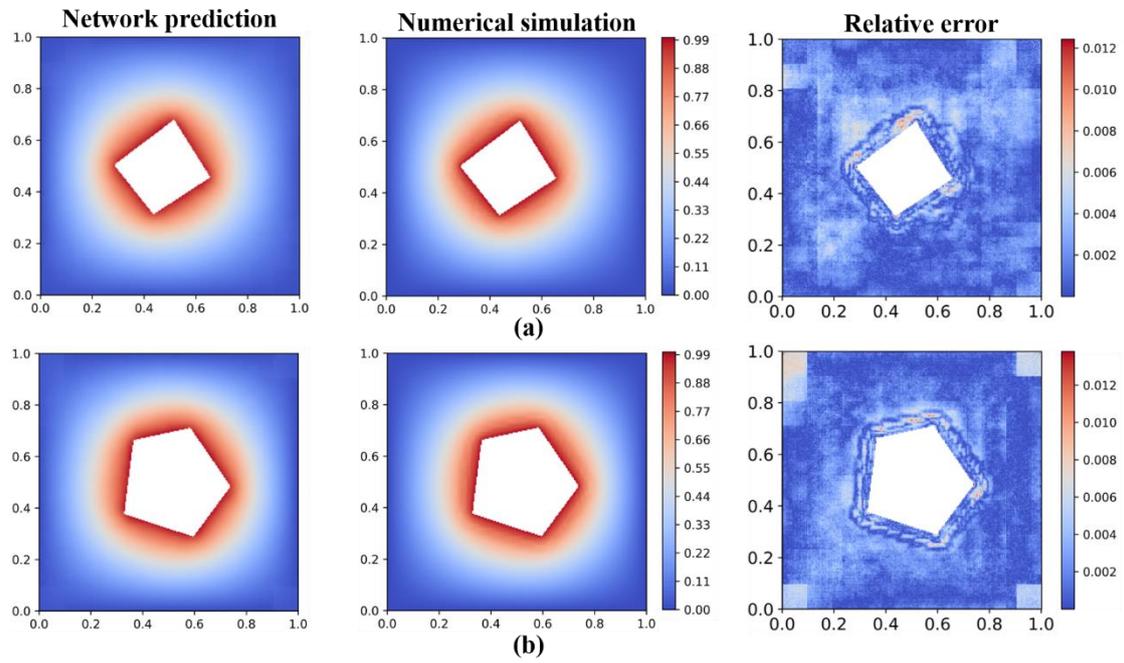

Fig. 7 (a) Temperature field and error distribution of a quadrangle case, (b) Temperature field and error distribution of a pentagon case. The first and second column shows the temperature field predicted by the network model and the numerical simulation (OpenFOAM). The third column shows the relative error distribution.

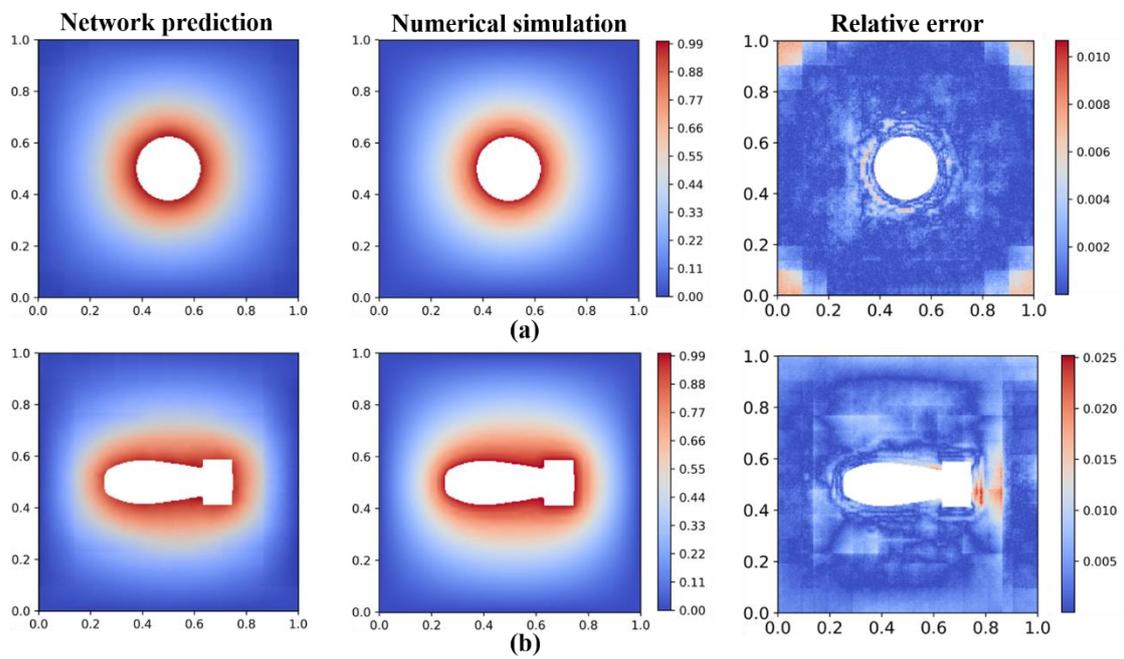

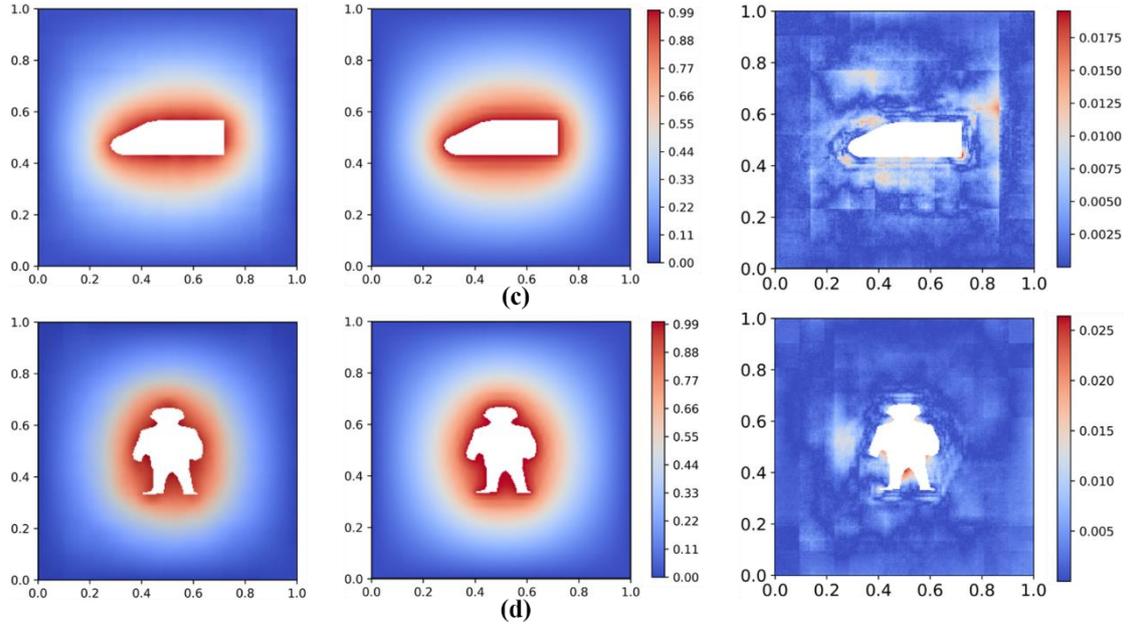

Fig. 8 Temperature field and relative error distribution of the validation cases with geometry of (a) circle, (b) bomb, (c) locomotive and (d) human.

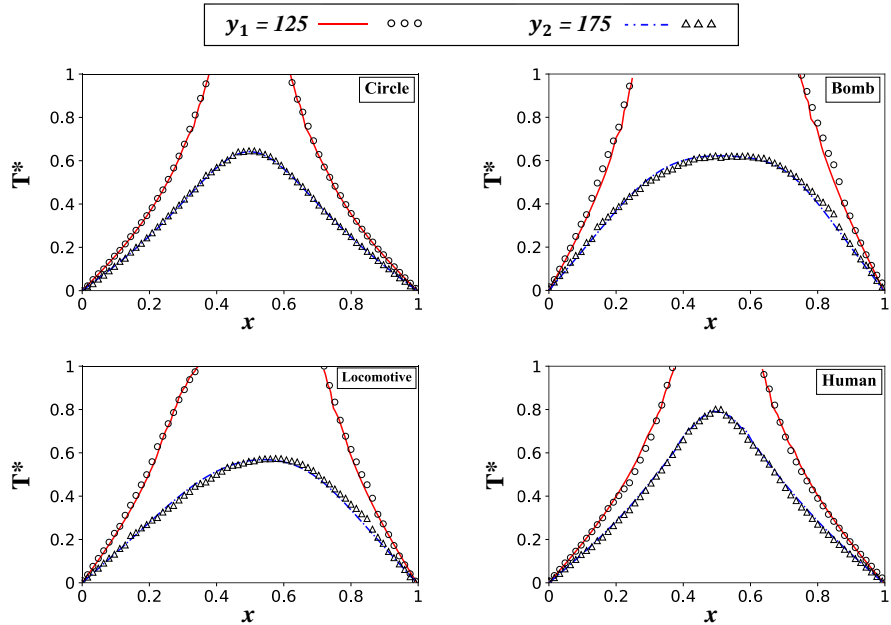

Fig. 9 Temperature profiles of the validation cases along x-direction at $y_1 = 125$; $y_2 = 175$, by the network model (lines) and numerical simulation (symbols).

Second, we study the time consumption of the prediction of the network model. As we know, GPU performs floating-point calculation faster than CPU, and the network model can be solved by GPU which further fasten the prediction. The time consumption for predicting the steady-state temperature field by the network model and the numerical simulation is shown in Table 4. Overall, the network model is faster than the numerical simulation for 3 to 4 order. As the geometry of the hot object becomes more complex, for the numerical simulation it requires finer mesh to converge the simulation, as a result the time consumption increases. However, the network model doesn't have any convergence problem and its prediction resolution is fixed, therefore the prediction time consumption keeps almost constant as the complexity of the problem geometry increases. Table 5 shows the average prediction accuracy on the different validation cases. It is surprised that the "most complex" case, the "Human" object, has higher accuracy than "Fish", "Fire", "Dog", even "Locomotive" and "Bomb".

Table 4 Time consumption of the network model prediction on GPU and numerical simulation on CPU using OpenFOAM.

| Geometry object | CNNs (s) | OpenFOAM (s) | Grid quantity | Speedup |
|---|---|---|---|---|
| Circle | 0.146144 | 530 | 5533 | 3626 |
| Locomotive | 0.13847 | 658 | 5694 | 4752 |
| Fire | 0.133502 | 671 | 5879 | 5026 |
| Bomb | 0.133615 | 909 | 11280 | 6803 |
| Fish | 0.133643 | 1288 | 13166 | 9637 |
| Human | 0.135947 | 1593 | 15187 | 11717 |
| Dog | 0.13464 | 1636 | 15812 | 12150 |

Table 5 Prediction accuracy of the network model on validation cases.

| Geometry object | Accuracy |
|---|---|
| Circle | 98.31% |
| Locomotive | 96.76% |
| Human | 97.27% |
| Fish | 93.92% |
| Bomb | 95.72% |
| Fire | 92.17% |
| Dog | 90.29% |

**3.2 Influence of the space distribution of the hot object**

In this part, the performance of the reduced-order model will be further investigated. Although, the network model has shown good prediction accuracy on validation dataset, the predicted objects above are intentionally chosen located in the center of the entire field. As mentioned in the section of data preparation, in general, it is more often that the object locates near the center of the studied domain, therefore the training dataset contains more cases in which the hot object locates near the center of the simulated domain. Here, we investigate the effect of the space distribution of the hot object on the performance of the network model. In detail, we change the distance between the center of the hot object (circle) and the center of the studied domain.

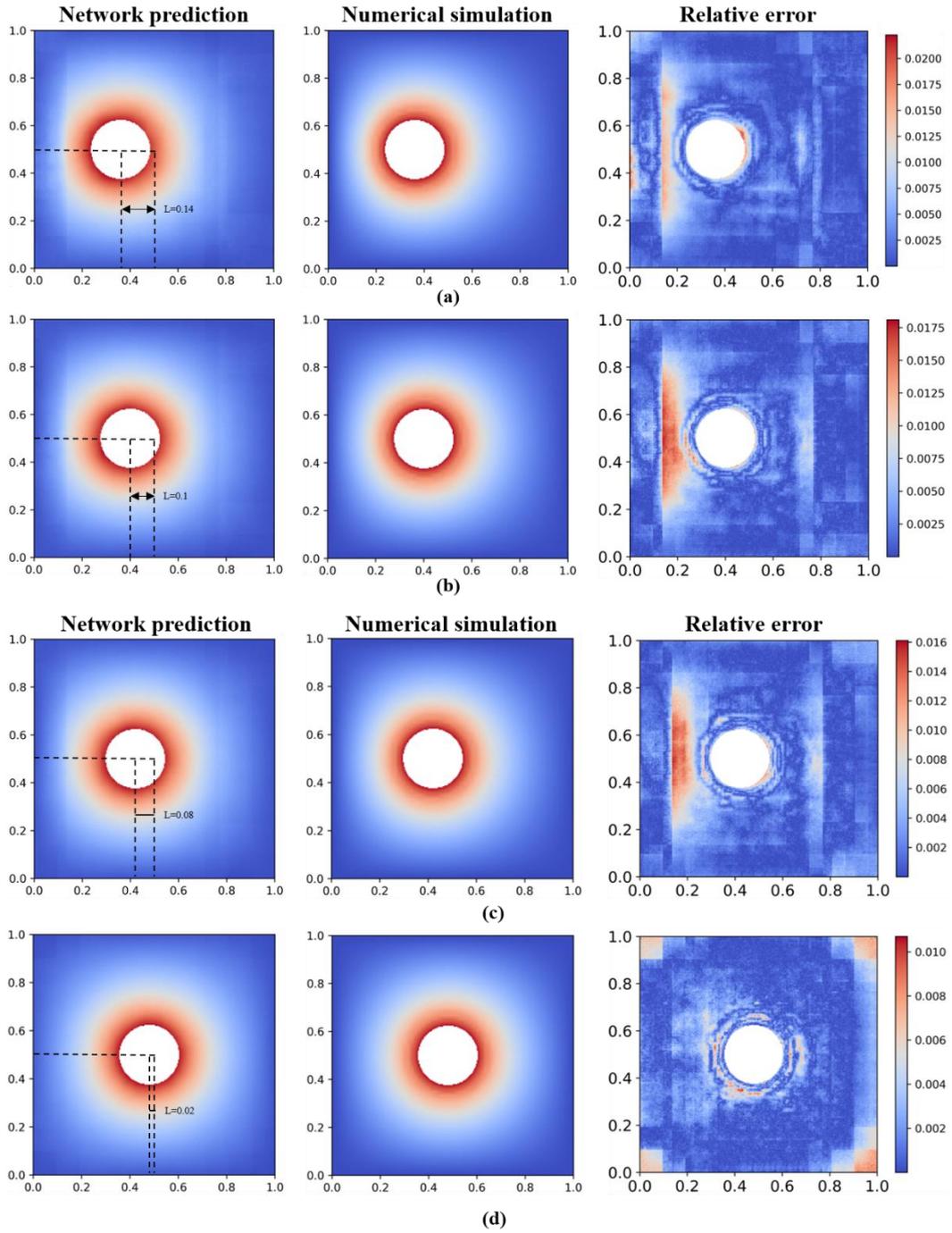

Fig. 10 Temperature distribution predicted by the network model and numerical simulation with different positions of the hot object, and the corresponding relative error. The distances between the center of the hot object and the center of the studied domain are (a) $L = 0.14$, (b) $L = 0.1$, (c) $L = 0.08$, (d) $L = 0.02$.

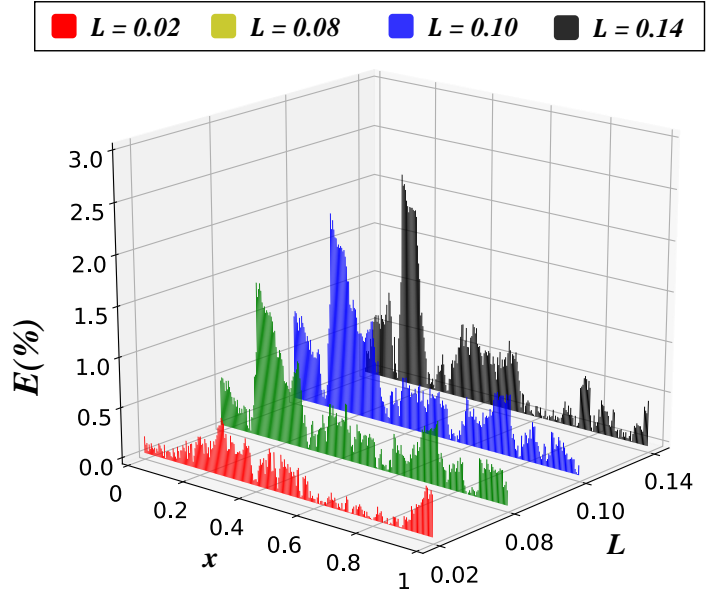

Fig. 11 Relative error profile along *x* direction (cross the center of the circle) with different *L*.

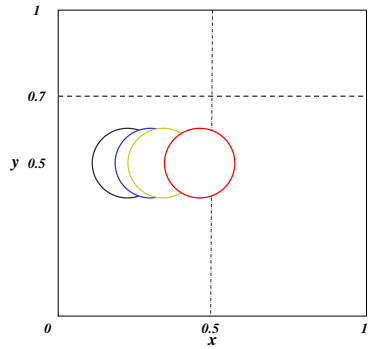

Fig. 12 Visualization of the position of the studied objects, where the Red circle denotes *L*=0.02, the yellow circle denotes *L*=0.08, the blue circle denotes *L*=0.1, and the black circle denotes *L*=0.14.

Fig. 10 and Fig. 11 show the prediction results by the network model and numerical simulation (OpenFOAM), and the corresponding relative error, as the position of the hot object changes. Fig. 12 visualizes the position variation of the studied objects. The above result shows that the maximum error is less than 3%; while with the increase of *L*, the error increases as consequence.

All the training cases only contains single hot object, here we also investigate performance of the network model on predicting the problems with multiple hot objects. Fig. 13 shows temperature fields of the cases with two circles predicted by the network model and numerical simulation, and the corresponding relative error. It can be observed that the network model still gives a reasonable prediction, and the maximum error is about 5%.

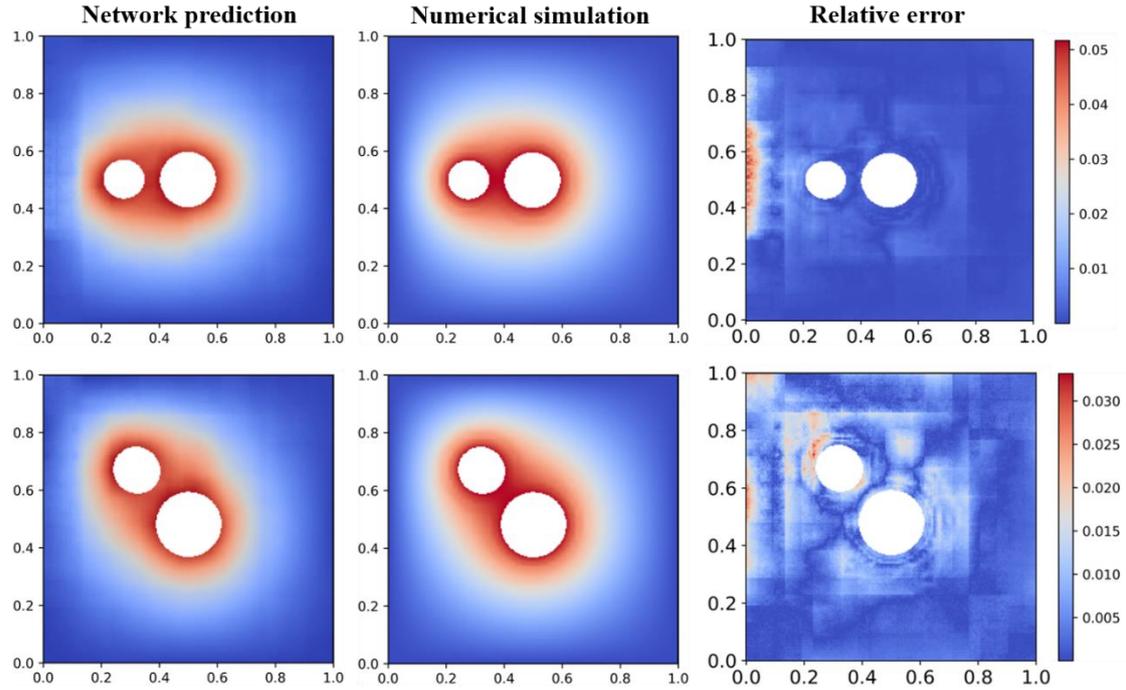

Fig. 13 Temperature fields of the cases with two circles predicted by the network model and numerical simulation, and the corresponding relative error.

## 4. Discussions

In principle, the reduced-order model is a data-driven method, and it can provide a fast prediction on the physical fields without solving partial differential equations (PDEs) iteratively and numerically. A fast and accurate reduced-order prediction model can be useful for the applications of production design and optimization, and the real-time simulation during some control problems. The results shown in last section indicate that the developed CNNs-based reduced-order model is able to accurately predict the steady-state temperature field even the geometry of the object is much complex than the training dataset. The strong extensionality on approximating highly non-linear system is one of the major advantages of the CNNs. Furthermore, the computational time can be negligible, as no iteration is need for the CNNs to finish a temperature field prediction.

However, there are also some drawbacks. As mentioned before, the training dataset contains more cases in which the hot object locates near the center of the simulated domain, therefore as the hot object is more far away from the center of the domain, the prediction accuracy decreases as consequence. One solution to this problem is to use more randomly distributed training dataset. Another possible drawback is that for a network model the physical property of the problem is difficult to interpret. In this paper, the model is designed and trained based on principle of obtaining the minimum prediction error, therefore it is still hard to establish an effective physical interpretation between the model structure and the physical laws of the problem, which is heat conduction in current paper. In fact, for guiding the design of the network, researchers have tried to find the similarity and connection between the architecture of the neural networks and the differential equations [39]–[42].

Even so, our results show that the developed network model can learn how to accurately predict temperature distribution of a steady-state heat conduction problem based on the training dataset. As the interdisciplinary use of the artificial neural networks is relatively new, we cannot claim this work has referred to full potential of the neural networks, and the discussed two drawbacks in last paragraph may can be also overcome in future. Therefore, it deserves paying more attention into the neural network-based modeling and see how it can better serve the physics or engineering community, such as heat transfer and fluid dynamics.

## 5. Conclusion

In this paper, a deep convolutional neural network (CNN) based reduced-order model (ROM) has been developed for fast prediction of temperature field of two-dimensional steady-state conduction, which is described by the Laplace's equation. In other words, the Laplace's equation is learned by the CNNs. Compared to the traditional ROM method, such as POD and DMD, the CNNs use a stack of simple functions (kernel) to approximate complicated differential equations, and can be seen as a discretization of the partial differential equations. Furthermore, in current work, the signed distance function is applied to represent the geometry of the problem. The dataset with simple polygons from triangle to dodecagon is used to train and test the model; and the performance of the network model is further validated and studied using much more complex geometries which have not been seen by the model during the training process.

The following conclusions are obtained from current study.

We adapt signed distance function (SDF) to represent the geometry of the problem. The SDF can carry a certain level of global information of the physical problem, which ensures the accuracy of the network model prediction. Compared to the traditional binary representation of the problem geometry, the network model using the SDF shows much better performance on the prediction accuracy.

The function, $\delta(s)$, is used to condition the loss function of the network mode prediction, see equation (8); and $\delta(s)$ like a mask matrix and has the same size as the studied domain. In detail, the loss inside the hot object will be ignored by applying $\delta(s)$. We compare the network model performance with or without applying the conditional function, and it shows that the neural network model performance much better with conditioning.

The most exciting observation of current work is that the dataset with simple geometries is feed into the network model as the training dataset, while the trained model is able to accurately predict the temperature distribution of the problems with much complex geometries, and the prediction speed of the network model is 3 to 4 order faster than numerical simulation using OpenFOAM. Using this approach, designers can easily generate numerous design alternatives in a negligible time during early design stages; or this approach can be used for real-time simulation during some control tasks.

## Acknowledgement

This work is supported by Natural Science Foundation of China No. 11802135, the Fundamental Research Funds for the Central Universities No. 30919011401.